\documentclass[amsmath,amssymb,showpacs,twocolumn,superscriptaddress,pr]{revtex4-1}
\usepackage{graphicx,bm,color,subfigure}

\begin{document}

\title{Revisitation of superconductivity in K$_2$Cr$_3$As$_3$ based on the six-band model}

\author{Li-Da Zhang}
\affiliation{School of Physics, Beijing Institute of Technology, Beijing 100081, China}

\author{Xianxin Wu}
\affiliation{Institute of Physics, Chinese Academy of Sciences, Beijing 100190, China}

\author{Heng Fan}
\affiliation{Institute of Physics, Chinese Academy of Sciences, Beijing 100190, China}
\affiliation{Collaborative Innovation Center of Quantum Matter, Beijing, China}

\author{Fan Yang}
\email{yangfan\_blg@bit.edu.cn}
\affiliation{School of Physics, Beijing Institute of Technology, Beijing 100081, China}

\author{Jiangping Hu}
\email{jphu@iphy.ac.cn}
\affiliation{Institute of Physics, Chinese Academy of Sciences, Beijing 100190, China}
\affiliation{Department of Physics, Purdue University, West Lafayette, Indiana 47907, USA}
\affiliation{Collaborative Innovation Center of Quantum Matter, Beijing, China}

\begin{abstract}
We investigate the pairing symmetry of the newly synthesized quasi-one-dimension K$_2$Cr$_3$As$_3$ superconductor based on the six-band model. We adopt standard random-phase-approximation to study the Hubbard-Hund model of the system. Our results confirm the conclusions obtained from our previous three-band model: the triplet $p_z$ and $f_{y^3-3x^2y}$ wave pairings serve as the leading pairing symmetries in the weak and strong Hund's rule coupling regimes, respectively. For physically realistic parameters, the triplet $p_z$-wave pairing driven by ferromagnetic fluctuations is the leading pairing symmetry of the system. The consistency between the results of these two models suggests that the obtained $p_z$-wave pairing symmetry is physical and model-independent.
\end{abstract}


\maketitle


The newly discovered chromium-based superconductivity has raised great research enthusiasm recently \cite{Luo,Kotegawa,KCrAs,RbCrAs,CsCrAs} due to possible unconventional superconductivity driven by the strong electron-electron correlations among the 3d electrons of the Cr atoms in this group \cite{Luo,Kotegawa,Zhaojun,NMRCrAs}. Among the Cr-based superconductors, a particularly intriguing family is the quasi-one-dimension (Q1D) A$_2$Cr$_3$As$_3$ \cite{KCrAs,RbCrAs,CsCrAs} (A=K, Rb, Cs) superconductors, whose crystal structure consists of alkali-metal-ion-separated [(Cr$_3$As$_3$)$^{2-}$]$_{\infty}$ double-walled subnanotubes. The $T_c$ of this family is up to 6.1K at ambient pressure \cite{KCrAs}. This superconducting family is striking because on the one hand Q1D superconductors are rare, and on the other hand, many experimental results on this family suggest that their superconducting pairing is unconventional, and probably is spin triplet \cite{KCrAs,RbCrAs,CsCrAs,NMRKCrAs,pendepth1,pendepth2,ZhongH}.

Different experiments performed on the A$_2$Cr$_3$As$_3$ family revealed that both their normal and superconducting states are abnormal and unconventional. In the normal state, the large specific heat coefficient \cite{KCrAs} suggests large effective mass enhancement possibly caused by strong electron-electron correlations; the linearly temperature-dependent resistivity \cite{KCrAs} and the abnormality in the NMR relaxation-rate \cite{NMRKCrAs} suggest possible Luttinger-liquid behavior. In the superconducting state, the experimental results of the London penetration depth in the mixed state\cite{pendepth1,pendepth2}, the NMR relaxation rate \cite{NMRKCrAs} and the specific heat \cite{RbCrAs} suggest line gap nodes in the system, which is a characteristic of unconventional SC. Moreover, the ferromagnetic fluctuations detected by the NMR \cite{NMRRbCrAs} might probably mediate triplet pairing in the system, as supported by the particularly high upper critical field largely exceeding the Pauli limit \cite{KCrAs,RbCrAs,CsCrAs}. Since triplet SC is rare and intriguing, this possibility is rather important.

An urgent important question for the A$_2$Cr$_3$As$_3$ superconductors family is: what's the pairing symmetry? Previously we have constructed a minimum three-band tight-binding model to answer this question \cite{3band}. In this model, the original lattice structure of the system is simplified as parallel placed chains of virtual Cr atoms, with each virtual Cr atom containing three 3d orbitals whose symmetries are $d_{z^2}$, $d_{xy}$ and $d_{x^2-y^2}$. Note that there are two types of inequivalent virtual Cr atoms staggered arrayed on a chain to form sublattices A and B. Taking into account the orbitals on both sublattices leads to a six-bands tight-banding (TB) model. In our previous study \cite{3band}, however, only the orbitals on sublattice B are considered, which leads to a three-bands TB model. The insight supporting this minimum three-band model comes from the fact that the Cr atoms on sublattice B contributes more to the density of state (DOS) near the Fermi level, and the three-band model correctly reproduces the Fermi surface (FS). Starting from this minimum three-band model, we have studied the pairing symmetry of the interacting system both in the weak-coupling and strong-coupling limits. In both limits, our calculations predict that the leading pairing symmetry of the system is the triplet $p_z$-wave one for realistic parameters. This triplet pairing is consistent with \cite{3bandphys} the particularly high upper critical field exceeding the Pauli limit \cite{KCrAs,RbCrAs,CsCrAs}, and whose symmetry-protected line gap-nodes on the $k_z=0$ plane is consistent with \cite{3bandphys} the different experiments introduced above \cite{RbCrAs,NMRKCrAs,pendepth1,pendepth2}. Besides, this triplet pairing can account for \cite{3bandphys} the anisotropic temperature-dependent superfluid density qualitatively.

In spite of its success, the three-band model has obvious drawbacks. On the aspect of energy dispersion, although the model reproduces the FS, its band structure slightly below the Fermi level is not quite consistent with that of the LDA band structure. Actually, in comparison with the LDA one \cite{Cao,Wu1,3band}, there are missing band and band crossing in the band structure of the three-band model just 0.1eV below the Fermi level. Such a distance is touchable by the interaction parameters. On the aspect of the orbital component, the three-band model only takes the contribution from the Cr atoms on sublattice B, which, however, only dominates that on sublattice A in the two Q1D bands. In the other three-dimensional (3D) band, the contributions from the Cr atoms on both sublattices are comparable. Bearing these drawbacks of the three-band model in mind, one cannot help ask oneself: Is the three-band model adequate in capturing the correct pairing symmetry? What if we start from a more complete band structure? Will the resulting pairing symmetry be model dependent? To settle this issue, we need to revisit the superconductivity of the system starting from the six-band TB model, whose band structure is consistent with the LDA one within a large energy scale.

In this paper, we study the pairing symmetry of K$_2$Cr$_3$As$_3$ from the six-band TB model with the Hubbard-Hund interaction terms. We adopt the random phase approximation (RPA) as an appropriate approach for the problem. The spin susceptibility of this six-band model still peaks at the $\Gamma$-point, suggesting ferromagnetic fluctuations. The phase diagram of the system is consistent with that obtained from the three-band model. In the parameter region with large Hubbard-$U$, the spin-density-wave (SDW) phase emerges. In the parameter region with small Hubbard-$U$, the on-site $f$-wave superconducting phase occurs for strong Hund's rule coupling, and the $p_z$-wave superconducting phase occurs for weak Hund's rule coupling. For realistic Hund's rule coupling in the system, the triplet $p_z$-wave pairing driven by ferromagnetic fluctuations should still be the leading pairing symmetry. The consistency between the results of these two models verifies that they can serve as a complete description of the band structure of K$_2$Cr$_3$As$_3$ to study the pairing symmetry, and that the obtained $p_z$-wave pairing symmetry is robust and model-independent.


The six-band TB Hamiltonian of the system can be expressed as \cite{3band}:
\begin{align}\label{hk}
h(\bm{k})=\left(\begin{array}{cccccc}
h_{11}^{AA}& h_{12}^{AA} & h_{13}^{AA} & h_{11}^{AB} & h_{12}^{AB} &h_{13}^{AB}
 \\
 & h_{22}^{AA} & h_{23}^{AA} & h_{21}^{AB} & h_{22}^{AB} &h_{23}^{AB} \\
 &               & h_{33}^{AA} & h_{31}^{AB} & h_{32}^{AB} &h_{33}^{AB} \\
  &               &              & h_{11}^{BB} & h_{12}^{BB} &h_{13}^{BB} \\
  &               &              &               & h_{22}^{BB} &h_{23}^{BB} \\
  &               &              &               &               &h_{33}^{BB} \\
 \end{array}\right).
\end{align}
Here $A$ and $B$ represent sublattices, the orbital indices $1$, $2$, and $3$ denote $d_{z^2}$, $d_{xy}$, and $d_{x^2-y^2}$ orbitals, respectively, and the unshown matrix elements can be obtained from the shown ones by the Hermicity of $h(\bm{k})$.

Introducing $x=\frac{\sqrt{3}}{2}{k_xa_0}$, $y=\frac{1}{2}k_ya_0$ and $z=\frac{1}{2}k_zc_0$, where the $a_0$ and $c_0$ are the lattice constants, the hoppings between $d_{z^2}$ and $d_{z^2}$ orbitals lead to the following matrix elements of $h(\bm{k})$:
\begin{align}\label{hop1}
h_{11}^{AA/BB}=&\epsilon_{1/3}+2\sum_{i=1}^{3}
s^{11}_{z,2i}\cos2iz+(s^{11}_y+2s^{11}_{yzz}\cos 2z)                   \nonumber\\
&\times(2\cos2y+4\cos x\cos y ),                                       \nonumber\\
h_{11}^{AB}=&2\sum_{i=1}^{4}s^{11}_{z,2i-1}\cos[(2i-1)z]               \nonumber\\
&+2s^{11}_{yz}\cos z (2\cos 2y+4\cos x \cos y),                        \nonumber\\
h_{11}^{BA}=&h_{11}^{AB,*}=h_{11}^{AB}.
\end{align}
Furthermore, the hoppings between $d_{z^2}$ and $(d_{xy},d_{x^2-y^2})$ orbitals lead to
\begin{align}\label{hop2}
h_{12}^{AA/BB}=&2is^{12}_{1y}\sin 2y
+2is^{12}_{1y}\sin y \cos x                                        \nonumber\\
&+2\sqrt{3}s^{12}_{2y}\sin y\sin x                                 \nonumber\\
h_{21}^{AA/BB}=&h_{12}^{AA/BB,*}                                   \nonumber\\
h_{13}^{AA/BB}=&2s^{12}_{2y}\cos 2y
-2i\sqrt{3}s^{12}_{1y}\cos y\sin x                                 \nonumber\\
&-2s^{12}_{2y}\cos y\cos x                                         \nonumber\\
h_{31}^{AA/BB}=&h_{13}^{AA/BB,*}                                   \nonumber\\
h_{12}^{AB}=&2\cos z (2is^{12}_{1yz}\sin 2y
+2is^{12}_{1yz}\sin y \cos x                                       \nonumber\\
&+2\sqrt{3}s^{12}_{2yz}\sin y \sin x)                              \nonumber\\
h_{21}^{AB}=&h_{12}^{AB,*},
h_{12}^{BA}=h_{12}^{AB},
h_{21}^{BA}=h_{12}^{AB,*}                                          \nonumber\\
h_{13}^{AB}=&2\cos z (2s^{12}_{2yz}\cos 2y
-2\sqrt{3}is^{12}_{1yz}\cos y \sin x                               \nonumber\\
&-2s^{12}_{2yz}\cos y \cos x)                                      \nonumber\\
h_{31}^{AB}=&h_{13}^{AB,*},
h_{13}^{BA}=h_{13}^{AB},
h_{31}^{BA}=h_{13}^{AB,*}.
\end{align}
Lastly, the hoppings between $(d_{xy},d_{x^2-y^2})$ and $(d_{xy},d_{x^2-y^2})$ orbitals lead to
\begin{align}\label{hop3}
h_{22}^{AA/BB}=&\epsilon_{2/4}+2s^{22}_{11y}\cos 2y
+(s^{22}_{11y}+3s^{22}_{22y})\cos x \cos y                             \nonumber\\
&+2\sum_{i=1}^{2}s^{22}_{z,2i}\cos2iz                                  \nonumber\\
h_{23}^{AA/BB}=&\sqrt{3}(s^{22}_{11y}
-s^{22}_{22y})\sin x \sin y +2is^{22}_{12y}\sin 2y                     \nonumber\\
&-4is^{22}_{12y}\cos x \sin y                                          \nonumber\\
h_{32}^{AA/BB}=&h_{23}^{AA/BB,*}                                       \nonumber\\
h_{33}^{AA/BB}=&\epsilon_{2/4}+2s^{22}_{22y}\cos 2y
+(3s^{22}_{11y}+s^{22}_{22y})\cos x \cos y                             \nonumber\\
&+2\sum_{i=1}^{2}s^{22}_{z,2i}\cos 2iz                                 \nonumber\\
h_{22}^{AB}=&2\sum_{i=1}^{3}s^{22}_{z,2i-1}\cos(2i-1)z
+2\cos z\left[2s^{22}_{11yz}\cos 2y \right.                            \nonumber\\
&\left.+(s^{22}_{11yz}+3s^{22}_{22yz})\cos x \cos y\right]             \nonumber\\
h_{22}^{BA}=&h_{22}^{AB,*}=h_{22}^{AB}                                 \nonumber\\
h_{23}^{AB}=&2\cos z\left[\sqrt{3}(s^{22}_{11yz}
-s^{22}_{22yz})\sin x \sin y \right.                                   \nonumber\\
&\left.+2is^{22}_{12yz}\sin 2y -4is^{22}_{12yz}\cos x \sin y\right]    \nonumber\\
h_{32}^{AB}=&h_{23}^{AB,*}, h_{23}^{BA}=h_{23}^{AB},
h_{32}^{BA}=h_{23}^{AB,*}                                              \nonumber\\
h_{33}^{AB}=&2\sum_{i=1}^{3}s^{22}_{z,2i-1}\cos(2i-1)z
+2\cos z\left[2s^{22}_{22yz}\cos 2y \right.                            \nonumber\\
&\left.+(3s^{22}_{11yz}+s^{22}_{22yz})\cos x \cos y\right]             \nonumber\\
h_{33}^{BA}=&h_{33}^{AB,*}=h_{33}^{AB}.
\end{align}
The hopping parameters in the above TB model can be obtained by least-square-root fitting of the band structure of the model to that of density functional theory (DFT). Here we refit these hopping parameters to highlight the Q1D feature of the system. The resulting values of these hopping parameters are listed in tables \ref{hopping1} and \ref{hopping2}.

\begin{table*}
\caption{The hopping parameters (in unit of eV) along $c$ axis to fit the DFT results in  the six-band TB model. The on-site energies are $\epsilon_1=1.8235$eV, $\epsilon_2=2.0763$eV, $\epsilon_3=1.9315$eV and $\epsilon_4=1.9761$eV.}
\label{hopping1}
\centering
\begin{tabular}{cccccccc}
$i$        & $z1$   & $z2$   & $z3$   & $z4$    & $z5$    & $z6$    & $z7$    \\
$s^{11}_i$ & 0.1402 & 0.1498 & 0.0367 & -0.0557 & -0.0175 & -0.0155 &  0.0053 \\
$s^{22}_i$ & 0.0243 & 0.1718 & 0.0004 & -0.0450 &  0      & -0.0022 & -0.0063
\end{tabular}
\end{table*}

\begin{table*}
\caption{The inplane hopping parameters (in unit of eV) to fit the DFT results in the six-band TB model.}
\label{hopping2}
\centering
\begin{tabular}{ccccccccc}
$i$          & $1y$   & $2y$   & $11y$   & $22y$   & $12y$  & $y$     & $yz$   & $yzz$  \\
$s^{12}_{i}$ & 0.0166 & 0.0103 &         &         &        &         &        &        \\
$s^{12}_{iz}$& 0.0093 & 0.0333 &         &         &        &         &        &        \\
$s^{22}_{i}$ &        &        &  0.0332 & -0.0177 & 0.0080 &         &        &        \\
$s^{22}_{iz}$&        &        & -0.0066 & -0.0078 & 0.0199 &         &        &        \\
$s^{11}_{i}$ &        &        &         &         &        & -0.0119 & 0.0015 & 0.0023
\end{tabular}
\end{table*}

\begin{figure}[htbp]
\centering
\includegraphics[width=0.48\textwidth]{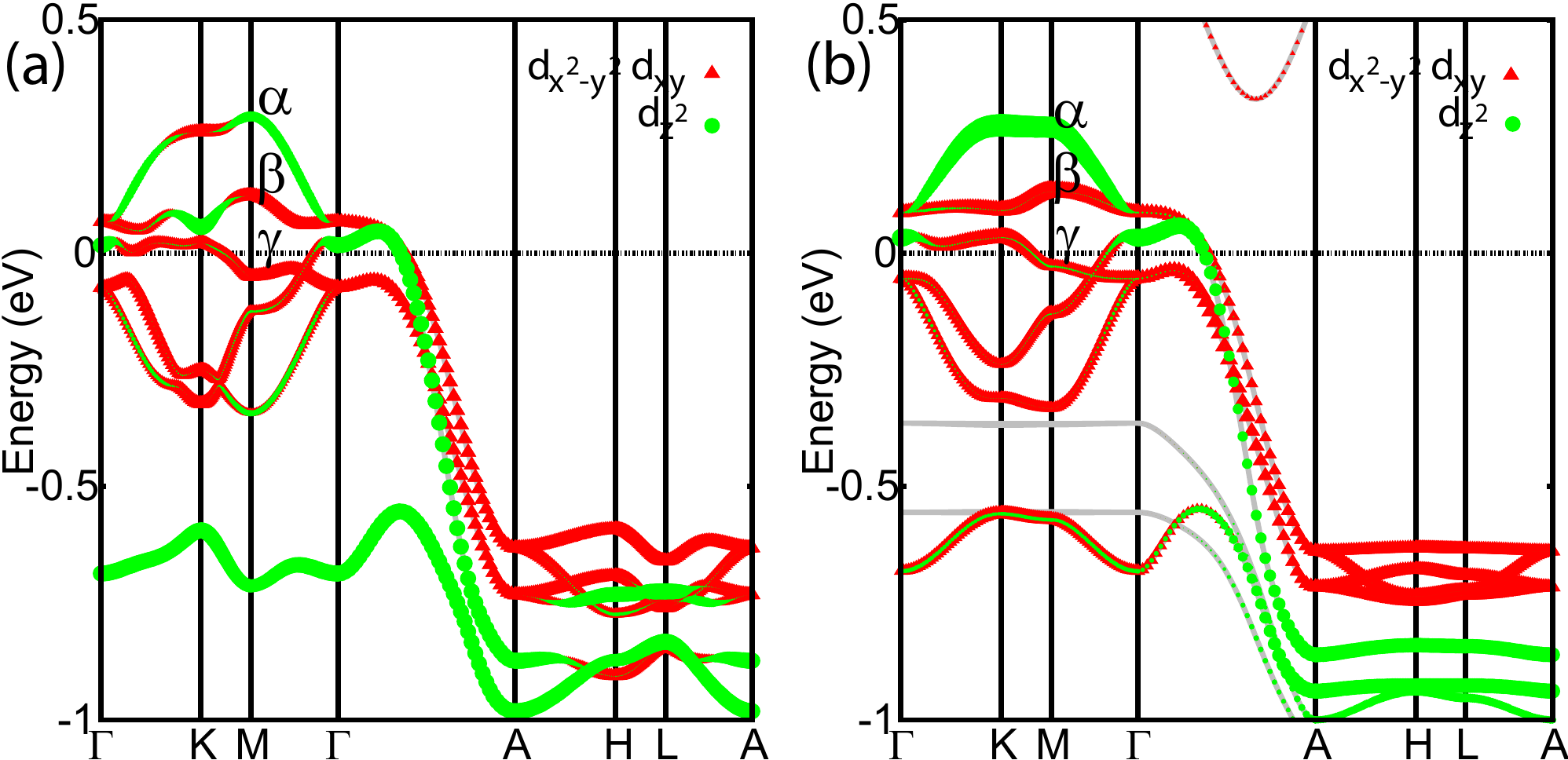}
\caption{The orbital characters of the bands in (a) the six-band TB model and (b) the DFT.}\label{band}
\end{figure}

\begin{figure}[htbp]
\centering
\includegraphics[width=0.48\textwidth]{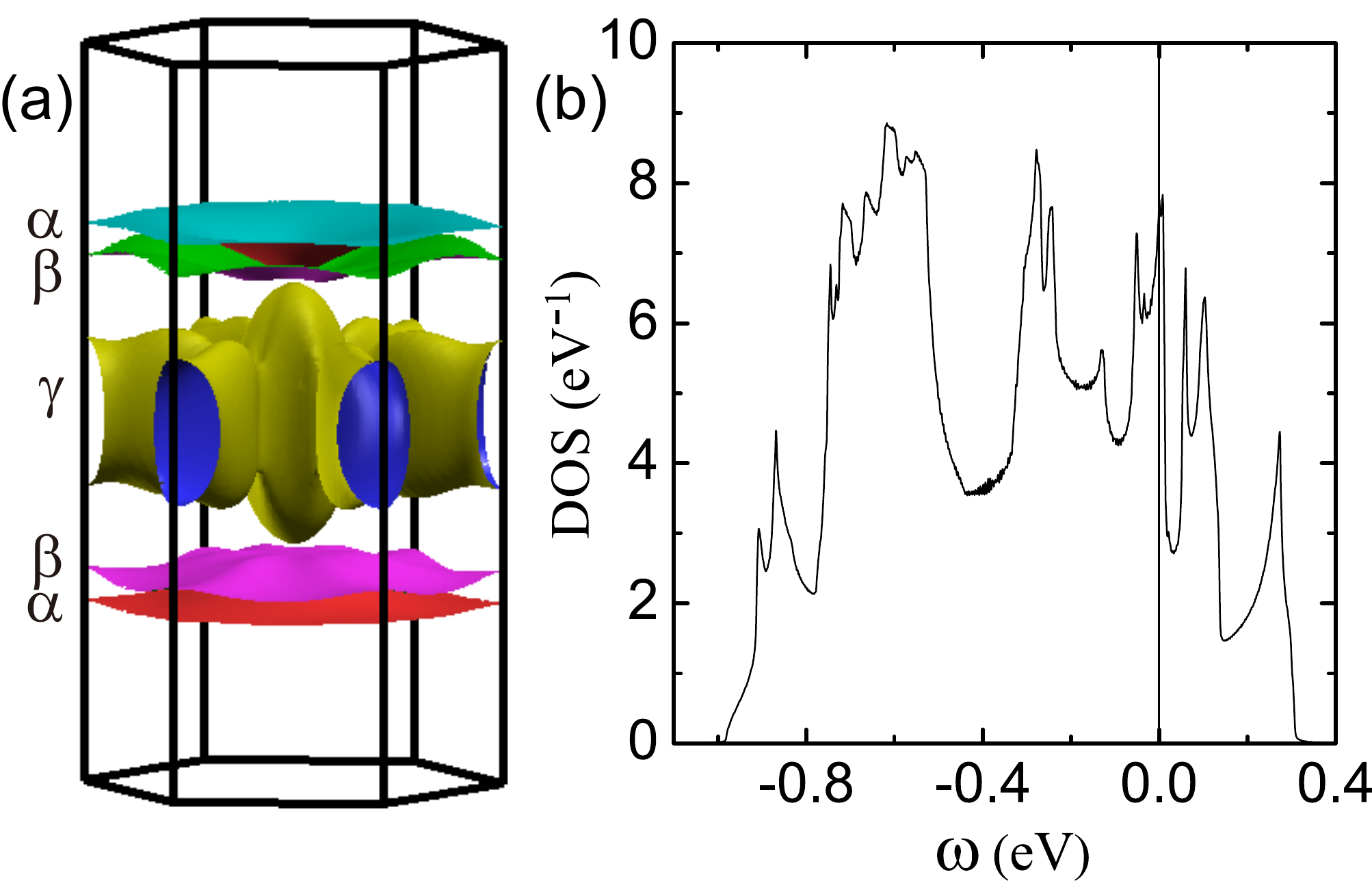}
\caption{(a) The Fermi surfaces and (b) the DOS for one spin specie of the six-band TB model.}\label{fsdos}
\end{figure}

The chemical potential of the six-band TB model is $\mu_c=2.3258$eV, which leads to a band filling of 10 electrons per unit cell. In figs. \ref{band}(a) and \ref{band}(b), we show the orbital characters of the bands of the six-band TB model and the DFT, respectively. From the orbital characters of the bands shown in fig. \ref{band}, we identify that the dominating orbital component on the $\alpha$-band is $d_{z^2}$, and those on both the $\beta$- and $\gamma$-bands are $d_{x^2-y^2}$ and $d_{xy}$, which is consistent with the result in the three-band model \cite{3band}. Among the six bands of the model, there are only three bands with higher energy intersecting with the Fermi level. These three bands correspond to those in the three-band model, and marked as $\alpha$, $\beta$, and $\gamma$ from high to low energy in fig. \ref{band}. In fig. \ref{fsdos}(a), we show the corresponding Fermi surfaces, which indicates that the $\alpha$- and $\beta$-bands are Q1D, and the $\gamma$-band is 3D. Due to the flatness of the two Q1D bands near the Fermi level, large density of states (DOS) is obtained there as shown in fig. \ref{fsdos}(b).

Following ref. \cite{3band}, we adopted the following Hubbard-Hund Hamiltonian in our calculations:
\begin{align}\label{model}
H=&H_{TB}+H_{I}\nonumber\\
H_{I}=&U\sum_{i\mu}n_{i\mu\uparrow}n_{i\mu\downarrow}+
V\sum_{i,\mu<\nu}n_{i\mu}n_{i\nu}+J_H\sum_{i,\mu<\nu}                   \nonumber\\
&\Big[\sum_{\sigma\sigma^{\prime}}c^{+}_{i\mu\sigma}c^{+}_{i\nu\sigma^{\prime}}
c_{i\mu\sigma^{\prime}}c_{i\nu\sigma}+(c^{+}_{i\mu\uparrow}c^{+}_{i\mu\downarrow}
c_{i\nu\downarrow}c_{i\nu\uparrow}+h.c.)\Big].
\end{align}
Here, the $U$, $V$, and $J_H$ terms denote the intra-orbital, inter-orbital Hubbard repulsion and the Hund's rule coupling as well as the pair hopping. Because general symmetry argument requires $U=V+2J_H$, we choose $U$ and the ratio $J_H/U$ to be independent tuning parameters and study the parameter dependence of the results.

\begin{figure}[htbp]
\centering
\includegraphics[width=0.4\textwidth]{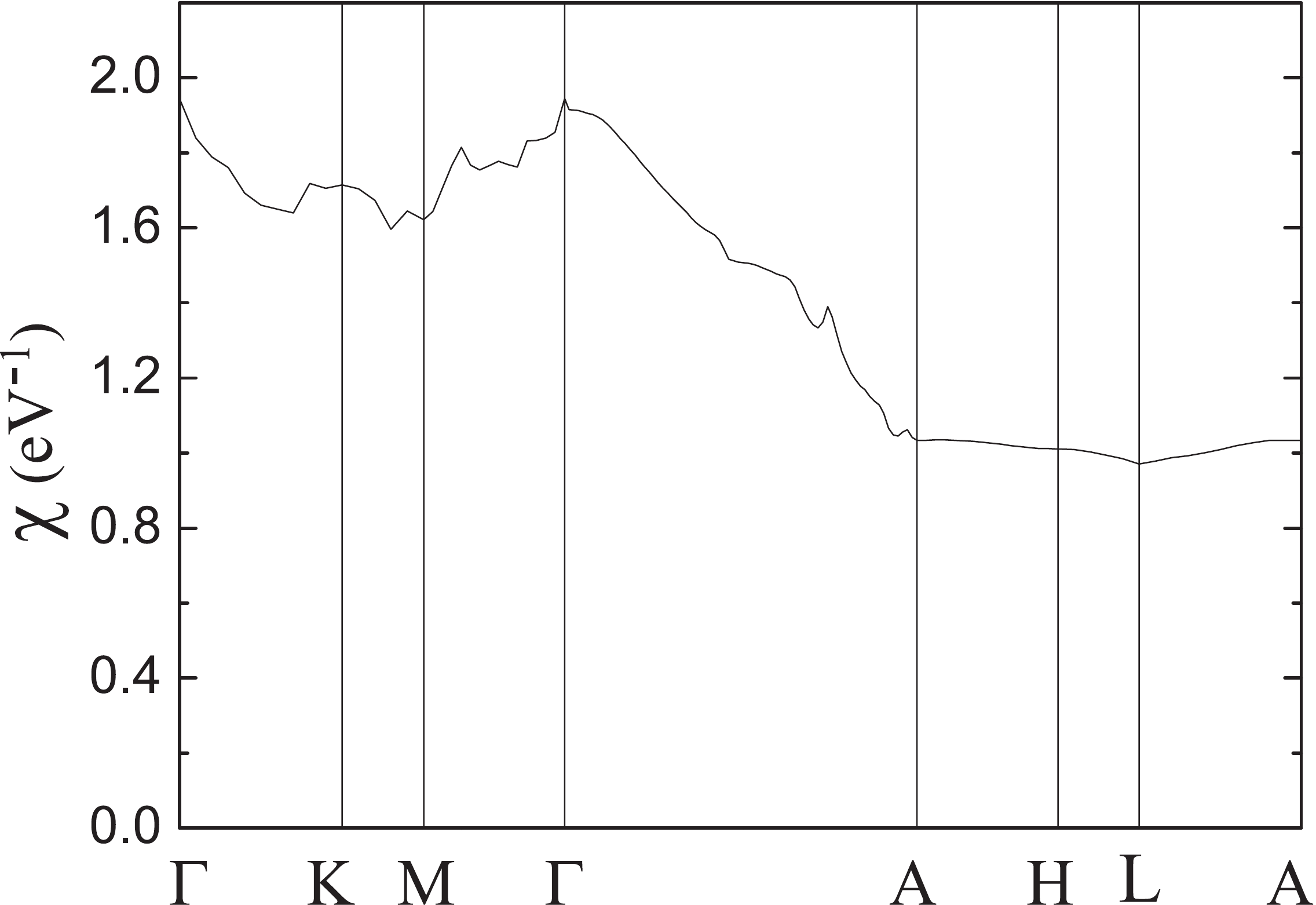}
\caption{The $\bm{k}$-dependence of the largest eigenvalue of susceptibility matrix $\chi^{(0)pp}_{ss}(\bm{k},i\omega_n=0)$ along the high-symmetry lines in the Brillouin Zone.}\label{chi}
\end{figure}

According to the standard multi-orbital RPA approach \cite{RPA1,RPA2,RPA3,Kuroki,Scalapino1,Scalapino2,Yang,Wu2014,Ma2014,Zhang2015} used in ref. \cite{3band}, for the non-interacting case where $U=V=J_H=0$, we can define the bare susceptibility $\chi^{(0)pq}_{st}(\bm{k},i\omega_n)$ in the $\bm{k}$-space with $p,q,s,t=1,2,...,6$ being the orbital indices and $i\omega_n$ being the Matsubara frequency. In fig. \ref{chi}, we show the $\bm{k}$-dependence of the largest eigenvalue of susceptibility matrix $\chi^{(0)pp}_{ss}(\bm{k},i\omega_n=0)$ in the Brillouin Zone. Clearly, the largest eigenvalue peaks at the $\Gamma$-point, which implies that the dominating intra-sublattice spin correlation is ferromagnetic. Here we note that the peak value of the largest eigenvalue is about 2eV$^{-1}$, which is much smaller than the one obtained from the three-band model \cite{3band}.

When interactions is turned on, we can further define the spin susceptibility $\chi^{(s)}$ and charge susceptibility $\chi^{(s)}$. In the RPA level, as the repulsive Hubbard-$U$ increase, the spin susceptibility is enhanced remarkably. When the Hubbard-$U$ is up to a critical strength $U_c$, which depends on the ratio $J_H/U$, the renormalized spin susceptibility diverges, which invalidates the RPA treatment and implies the formation of the SDW phase.


When the interaction strength $U<U_c$, there are short-ranged spin or charge fluctuations in the system. Through exchanging these fluctuations between a Cooper pair, exotic superconducting states would emerge in the system. Corresponding to the $D_{3h}$ point group of K$_2$Cr$_3$As$_3$, there are ten possible superconducting pairing symmetries  \cite{3band}. Based on the six-band TB model, our calculations performing in different parameter regimes identify the leading pairing symmetries of the system as the triplet $p_z$-wave symmetry and the triplet $f_{y^3-3x^2y}$-wave one in the weak and strong Hund's coupling regimes, respectively. Because the orbitals in the present six-band model are the molecule orbitals that are distributed on the original six Cr atoms in a unit cell of the system, the Hund's coupling should be weak, and thus the $p_z$-wave pairing should be the leading pairing symmetry of the system. This result is qualitatively the same as the previous one obtained from the three-band model.

\begin{figure*}[htbp]
\centering
\includegraphics[width=0.9\textwidth]{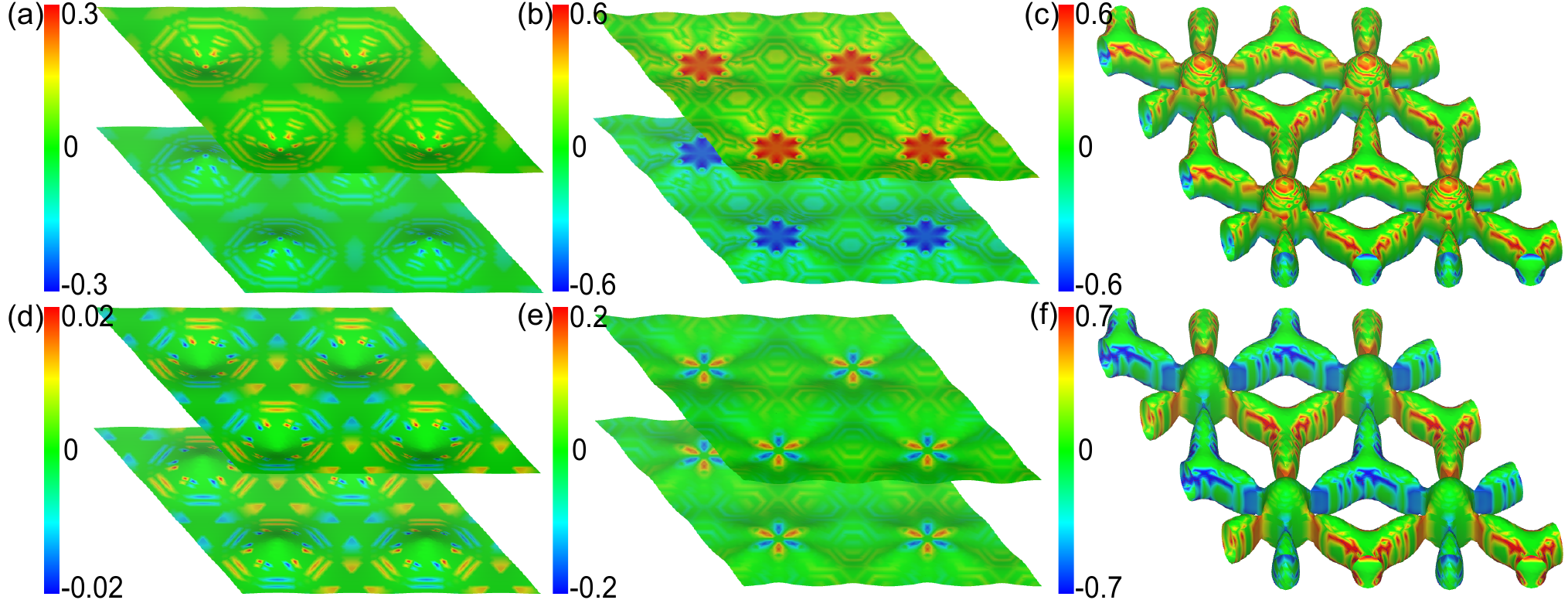}
\caption{(a), (b), and (c) are the relative gap functions of the $p_z$-wave pairing on FSs $\alpha$, $\beta$, and $\gamma$, respectively. (d), (e), and (f) are those of the $f_{y^3-3x^2y}$-wave pairing on the FSs $\alpha$, $\beta$, and $\gamma$, respectively. In (a), (b), and (c), we take $U=0.15$eV and $J_H=0.1U$. In (d), (e), and (f), we take $U=0.15$eV and $J_H=0.4U$.}\label{gap}
\end{figure*}

In fig. \ref{gap}(a), \ref{gap}(b), and \ref{gap}(c), we show the relative gap function of the leading $p_z$-wave pairing on the FSs of the six-band model. This triplet pairing is mirror-reflection odd about the $k_z=0$ plane, and thus has gap nodes on the intersecting line of the $k_z=0$ plane and the 3D FS $\gamma$. In the region $k_z>0$ or $k_z<0$ on the FSs, the gap function of the $p_z$-wave pairing doesn't change sign, and thus has no extra gap nodes. In fig. \ref{gap}(d), \ref{gap}(e), and \ref{gap}(f), we show the relative gap function of the leading $f$-wave pairing on the FSs of the six-band model. This triplet pairing state is mirror-reflection even about the $k_z=0$ plane. Its gap function changes sign over every 60$^{o}$ degree rotation about the $z$-axis, which leads to line gap nodes in the direction of $k_y=0,\pm\sqrt{3}k_x$.

From the relative amplitude of the gap functions of the leading $p_z$-wave pairing on different FSs shown in fig. \ref{gap}(a), \ref{gap}(b), and \ref{gap}(c), we find that the superconducting pairings on FSs $\beta$ and $\gamma$ are obvious stronger than that on FS $\alpha$. From the relative magnitude of the gap functions of the leading $f$-wave pairing on different FSs shown in fig. \ref{gap}(d), \ref{gap}(e), and \ref{gap}(f), we find that the pairings on FS $\gamma$ are even stronger than that on FS $\beta$, not to mention that on FS $\alpha$. This result is obviously different from the one obtained from the three-band model that the superconducting pairings on FS $\beta$ is always the strongest \cite{3band}. Here we emphasize that, in spite of this difference, the results in both the three-band and six-band models indicates that the pairing mainly takes place in the $d_{x^2-y^2}$ and $d_{xy}$ orbitals since these two orbitals dominate on both $\beta$- and $\gamma$-bands as above-mentioned.

\begin{figure}[htbp]
\centering
\includegraphics[width=0.4\textwidth]{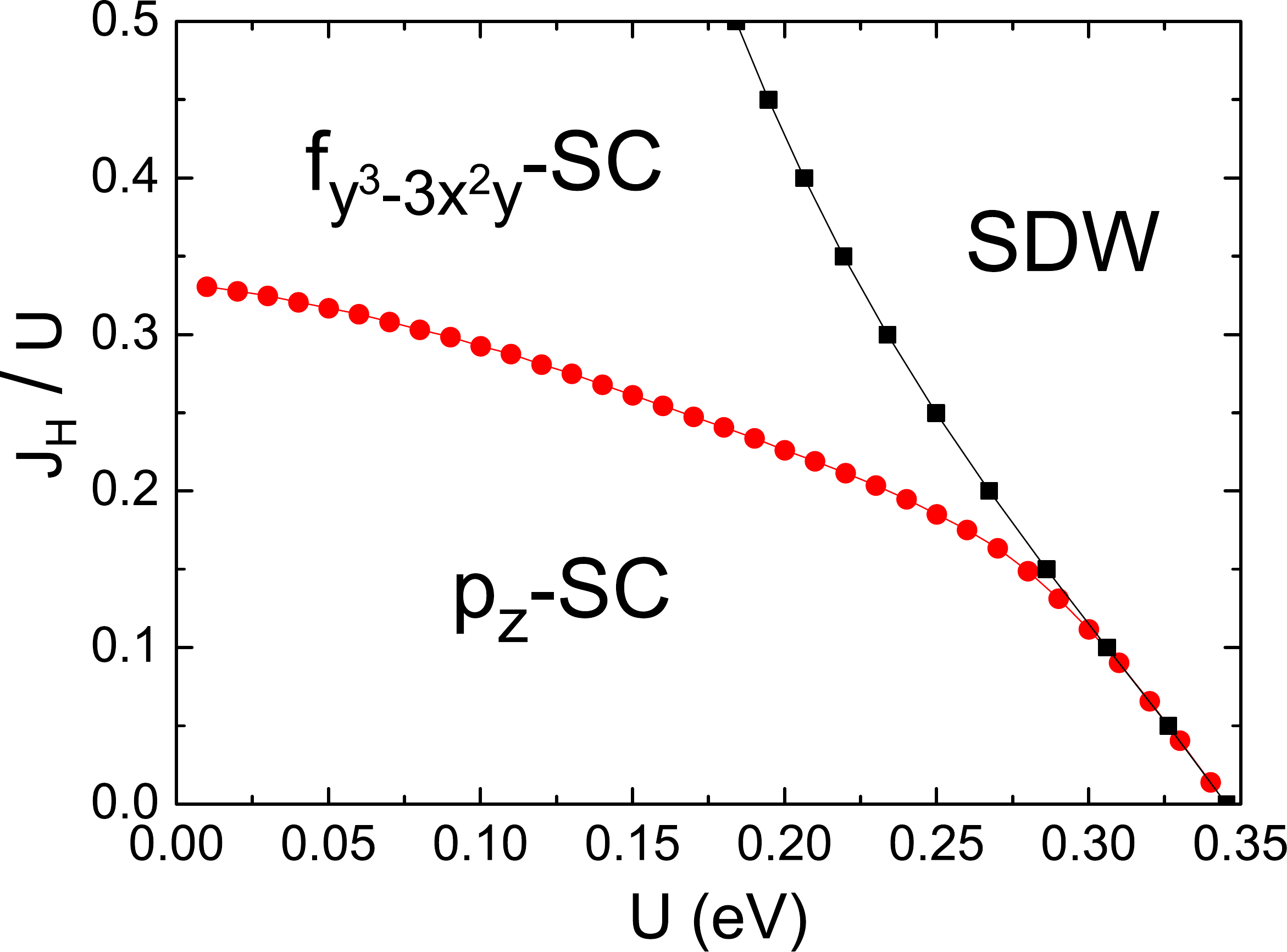}
\caption{The ground state phase diagram in the $U-J_H/U$ plane.}\label{phase}
\end{figure}

In fig. \ref{phase}, we show the ground state phase diagram on the $U-J_H/U$ plane. For $U>U_c$, which is $J_H/U$-dependent and ranges from $0.18$eV to $0.35$eV, the intra-sublattice ferromagnetic SDW state emerges; for $U<U_c$, the triplet $p_z$ and $f_{y^3-3x^2y}$ wave pairing states occur below and above a $U$-dependent critical value of $J_H/U$, respectively. Here we note that the $U_c$ obtained here is much larger than the one in the three-band model \cite{3band}, i.e., the superconducting phases obtained here extend to the parameter region with the larger $U$. The enhancement of $U_c$ can be understood from the suppression of the bare susceptibility $\chi^{(0)}$ comparing with the one in the three-band model, which requires a much larger $U_c$ to make the renormalized spin susceptibility $\chi^{(s)}$ diverge.

As for the value of the Hubbard-$U$, we note that, while the $d$ orbitals of the virtual atom in the three-band TB model of the system is the superposition of the original $d$ orbitals of the Cr atom on both sublattices A and B, the $d$ orbitals of the virtual atom in the present six-band model is the superposition of those on only one sublattice. Therefore, the $d$ orbitals in the present six-band model should be more localized than those in the three-band model, which implies that the effective $U$ in the present six-band model should be larger than the one in the three-band model.

\begin{figure}[htbp]
\centering
\includegraphics[width=0.48\textwidth]{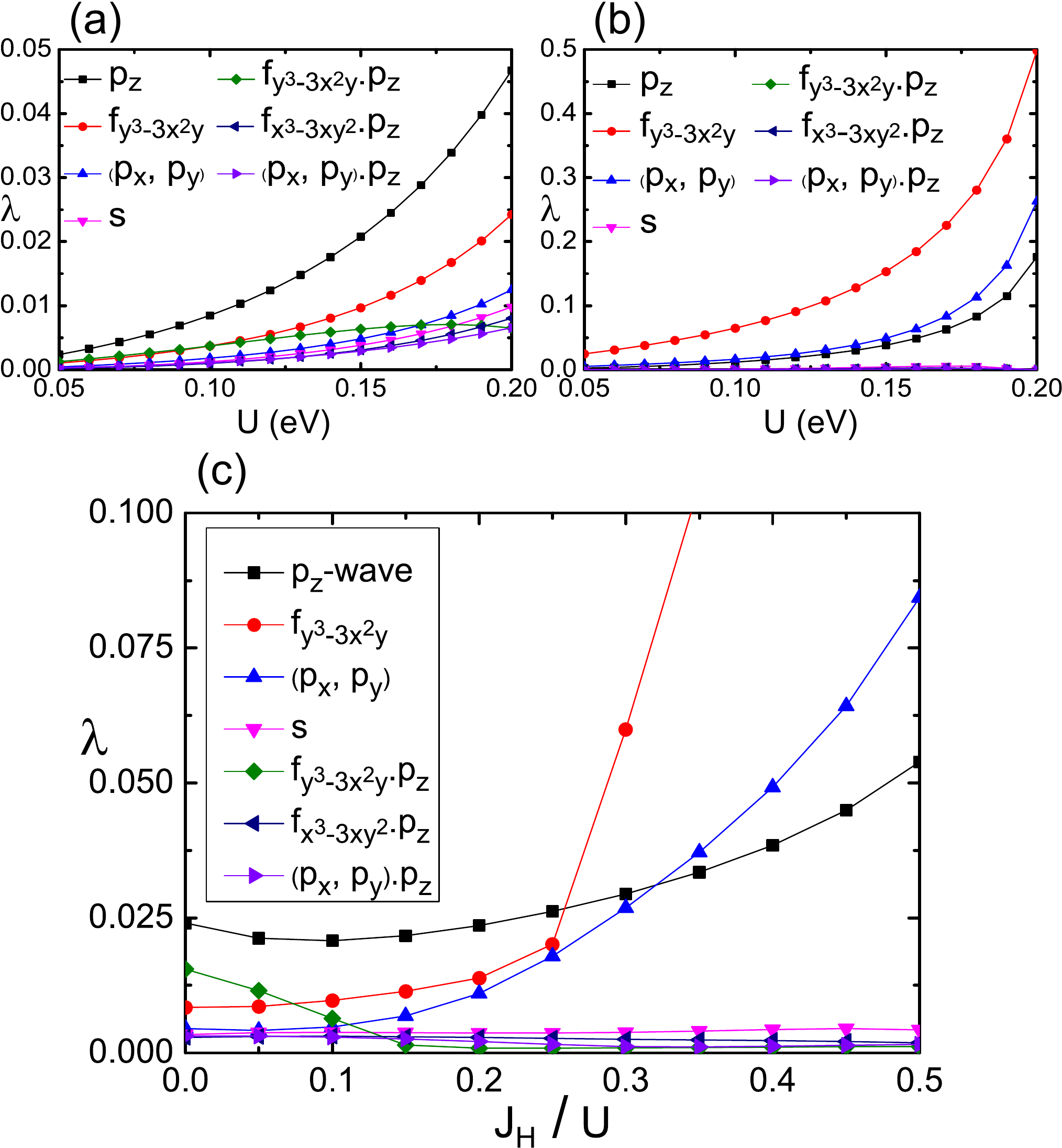}
\caption{The $U$-dependence of the largest eigenvalues $\lambda$ of seven stronger pairing symmetries for (a) $J_H=0.1U$ and (b) $J_H=0.4U$. (c) The $J_H/U$-dependence of the largest eigenvalues $\lambda$ of these pairing symmetries for fixed $U=0.15$eV.}\label{u}
\end{figure}

In fig. \ref{u}, we show the $U$-dependence of the largest eigenvalue $\lambda$ for the seven stronger pairing symmetries. Clearly, all these eigenvalues are enhanced promptly as $U$ increases and would diverge when $U$ tends to $U_c$. For $J_H=0.1U$, the $p_z$-wave is the leading pairing symmetry and dominates other symmetries as shown in fig. \ref{u}(a); for $J_H=0.4U$, the $f_{y^3-3x^2y}$-wave becomes the leading one, with the $(p_x,p_y)$ and $p_z$ wave pairing to be close candidates as shown in fig. \ref{u}(b). In fig. \ref{u} (c), we show the $J_H/U$-dependence of these eigenvalues for fixed $U=0.15$eV. Obviously, the triplet $p_z$ and $f_{y^3-3x^2y}$ symmetries dominate other symmetries at all relevant values of ratio $J_H/U$.


In conclusion, we have performed RPA study of the superconducting pairing symmetry of K$_2$Cr$_3$As$_3$ starting from the six-band TB model. Our calculations confirm the results of our previous study based on the three-band model: the triplet $p_z$ and $f_{y^3-3x^2y}$ wave pairings serve as the leading pairing symmetries in the weak and strong Hund's coupling regimes respectively, and the former wins over the latter for real material. The consistency between the results of these two models suggests that the three-band and six-band models are sufficient to capture the correct pairing symmetry of the system. While the obtained $p_z$-wave pairing symmetry is robust and model-independent, there is slight difference between the form of their pairing gap functions. The difference lies in that, while the gap amplitude for the three-band model mainly distributes on one of the Q1D FS, that for the present six-band model mainly distributes on the three-dimensional FS as well as this Q1D one. Such a property is left for experimental verification.

\textbf{Acknowledgments}: This work is supported in part by MOST of China (Grant Nos. 2011CBA00100, 2012CB821400, 2015CB921300), the NSFC (Grant Nos. 11190020, 11274041, 11334012, 91221303) and ``Strategic Priority Research Program (B)" of the Chinese Academy of Sciences (Grant No. XDB07020200). F.Y is also supported by the NCET program under Grant No. NCET-12-0038.


\begin{thebibliography}{*}

\bibitem{Luo}
W. Wu \emph{et al}, Nature Communications \textbf{5}, 5508 (2014).


\bibitem{Kotegawa}
H. Kotegawa, S. Nakahara, H. Tou, and H. Sugawara, J. Phys. Soc. Jpn. \textbf{83}, 093702 (2014).


\bibitem{KCrAs}
J.-K. Bao \emph{et al}, Phys. Rev. X \textbf{5}, 011013 (2015).


\bibitem{RbCrAs}
Z.-T.Tang \emph{et al}, Phys. Rev. B \textbf{91}, 020506(R) (2015).


\bibitem{CsCrAs}
Tang Z.-T. \emph{et al}, Science China Materials \textbf{58}, 16 (2015).


\bibitem{Zhaojun}
Shen Y. \emph{et al} arXiv:1409.6615 preprint, 2014.


\bibitem{NMRCrAs}
H. Kotegawa \emph{et al}, Phys. Rev. Lett. \textbf{114}, 117002 (2015).


\bibitem{NMRKCrAs}
H.-Z. Zhi, T. Imai, F.-L. Ning, J.-K. Bao, and G.-H. Cao, Phys. Rev. Lett. \textbf{114}, 147004 (2015).


\bibitem{pendepth1}
G.-M. Pang \emph{et al}, Phys. Rev. B \textbf{91}, 220502(R) (2015).


\bibitem{pendepth2}
G.-M. Pang \emph{et al}, Journal of Magnetism and Magnetic Materials, http://dx.doi.org/10.1016/j.jmmm.2015.08.093 (2015).


\bibitem{ZhongH}
H. Zhong, X.-Y. Feng, H. Chen, J. Dai, arXiv:1503.08965 preprint, 2015.


\bibitem{NMRRbCrAs}
J. Yang, Z.-T. Tang, G.-H. Cao, and G.-q. Zheng, Phys. Rev.  Lett. \textbf{115}, 147002 (2015).


\bibitem{Cao}
H. Jiang, G. Cao, and C. Cao, arXiv:1412.1309 preprint, 2014.


\bibitem{Wu1}
X. Wu, C. Le, J. Yuan, H. Fan, and J. Hu, Chin. Phys. Lett. \textbf{32}, 057401 (2015).


\bibitem{3band}
X. Wu, F. Yang, C. Le, H. Fan, and J. Hu, Phys. Rev. B \textbf{92}, 104511 (2015).


\bibitem{3bandphys}
X. Wu, F. Yang, S. Qin, H. Fan, and J. Hu, arXiv:1507.07451 preprint, 2015.


\bibitem{RPA1}
T. Takimoto, T. Hotta, and K. Ueda, Phys. Rev. B \textbf{69}, 104504 (2004).


\bibitem{RPA2}
K. Yada and H. Kontani, J. Phys. Soc. Jpn. \textbf{74}, 2161 (2005).


\bibitem{RPA3}
K. Kubo, Phys. Rev. B \textbf{75}, 224509 (2007).


\bibitem{Kuroki}
K. Kuroki \emph{et al}, Phys. Rev. Lett. \textbf{101}, 087004 (2008).


\bibitem{Scalapino1}
S. Graser, T. A. Maier, P. J. Hirschfeld, and D. J., Scalapino, New J. Phys. \textbf{11}, 025016 (2009).


\bibitem{Scalapino2}
T. A. Maier, S. Graser, P. J. Hirschfeld, and D. J. Scalapino, Phys. Rev. B, \textbf{83}, 100515(R) (2011).


\bibitem{Yang}
F. Liu, C.-C. Liu, K. Wu, F. Yang, and Y. Yao, Phys. Rev. Lett. \textbf{111}, 066804 (2013).


\bibitem{Wu2014}
X. Wu, J. Yuan, Y. Liang, H. Fan, and J. Hu, Europhys. Lett. \textbf{108}, 27006 (2014).


\bibitem{Ma2014}
T. Ma, F. Yang, H. Yao, and H. Lin, Phys. Rev. B \textbf{90}, 245114 (2014).


\bibitem{Zhang2015}
L.-D. Zhang, F. Yang, and Y. Yao, Sci. Rep. \textbf{5}, 8203 (2015).

\end{thebibliography}
\end{document}